\documentclass[fleqn,10pt]{wlscirep}
\title{Strong-field ionization of clusters using two-cycle pulses at 1.8~$\mu$m}

\author[1,*]{Bernd Sch\"utte}
\author[1]{Peng Ye}
\author[2]{Serguei Patchkovskii}
\author[1]{Dane R. Austin}
\author[1]{Christian Brahms}
\author[1]{Christian Str\"uber}
\author[1]{Tobias Witting}
\author[1,2]{Misha Yu. Ivanov}
\author[1]{John W. G. Tisch}
\author[1,+]{Jon P. Marangos}

\affil[1]{Department of Physics, Imperial College London, South Kensington Campus, London SW7
2AZ, UK.}
\affil[2]{Max-Born-Institut, Max-Born-Strasse 2A, 12489 Berlin, Germany.}
\affil[*]{bschutte@imperial.ac.uk}
\affil[+]{j.marangos@imperial.ac.uk}

\begin{abstract}
The interaction of intense laser pulses with nano-scale particles leads to the production of high-energy electrons, ions, neutral atoms, neutrons and photons. Up to now, investigations have focused on near-infrared to X-ray laser pulses consisting of many optical cycles. Here we study strong-field ionization of rare-gas clusters ($10^3$ to $10^5$ atoms) using two-cycle 1.8~$\mu$m laser pulses to access a new interaction regime in the limit where the electron dynamics are dominated by the laser field and the cluster atoms do not have time to move significantly. The emission of fast electrons with kinetic energies exceeding 3~keV is observed using laser pulses with a wavelength of 1.8~$\mu$m and an intensity of $1\times 10^{15}$~W/cm$^2$, whereas only electrons below 500~eV are observed at 800~nm using a similar intensity and pulse duration. Fast electrons are preferentially emitted along the laser polarization direction, showing that they are driven out from the cluster by the laser field. In addition to direct electron emission, an electron rescattering plateau is observed. Scaling to even longer wavelengths is expected to result in a highly directional current of energetic electrons on a few-femtosecond timescale.
\end{abstract}

\begin{document}
\flushbottom
\maketitle

\section*{Introduction}

In the past few years, ultrashort infrared (IR) laser pulses with wavelengths $>1$~$\mu$m have become an important tool for the investigation of strong-field physics in atoms, molecules and solids~\cite{colosimo08,blaga09,popmintchev12,weisshaupt14,schubert14,hohenleutner15,vampa15,huismans11}. The use of these laser pulses is often motivated by the large ponderomotive energies that electrons can acquire, which scale as the square of the wavelength. As a consequence, electrons can be efficiently accelerated up to high kinetic energies, which has been exploited for the production of high-harmonic generation (HHG) pulses in the keV range~\cite{popmintchev12} and for the development of bright hard x-ray sources~\cite{weisshaupt14}. Furthermore, the use of strong IR laser fields has led to an advancement of HHG schemes in solids~\cite{schubert14,hohenleutner15,vampa15} and to the development of time-resolved photoelectron holography~\cite{huismans11}. 


Sub-wavelength nanoscale particles represent a distinct class of targets, and were found to exhibit a fundamentally different behaviour in comparison to atoms, molecules and solids, when exposed to strong laser fields. While the particles themselves have a near-solid density, the distances between the particles are large, meaning that energy cannot be dissipated into the environment, as is the case for solids. The ionization of nanoscale clusters by intense laser pulses has been extensively studied in different wavelength regimes during the past two decades. Up to now, the largest number of experiments has been performed using laser pulses $<1$~$\mu$m~\cite{mcpherson94,ditmire95,snyder96,shao96,ditmire97a,ditmire99,doppner10,skopalova10,krishnan11,rajeev13,schutte15b,schutte15c,schutte16a}. Particularly intriguing results in this wavelength range were the emission of bright X-ray radiation~\cite{mcpherson94,ditmire95}, highly charged ions~\cite{snyder96}, fast ions~\cite{ditmire97a} and neutrals~\cite{rajeev13} with MeV kinetic energies, as well as the observation of correlated electronic decay~\cite{schutte15c}. The efficient absorption of near-infrared (NIR, which in this manuscript only refers to the wavelength regime around 800~nm) laser energy in clusters creates extreme conditions, which even resulted in the observation of nuclear fusion~\cite{ditmire99}. With the advent of free-electron laser and high-flux HHG sources, intense laser-cluster experiments have been extended to shorter wavelengths in the extreme-ultraviolet (XUV)~\cite{wabnitz02,laarmann05,bostedt08,iwayama10,bostedt12,murphy08,schutte14a,schutte14b} and X-ray regimes~\cite{gorkhover12,tachinaba15,gorkhover16}. 

Ionization of clusters by intense IR laser pulses $>1$~$\mu$m allows one to exploit both the high ponderomotive energies that electrons can acquire in the laser field and the efficient absorption of laser energy by nanoscale particles. This provides potential application as a source of high-energy electron, ion and photon emission. Up to now, IR strong-field ionization dynamics of clusters using laser wavelengths $>1$~$\mu$m have been largely unexplored, in contrast to the wavelength regimes from the NIR to the X-ray range. In Ref.~\cite{negro14}, XUV fluorescence from molecular clusters was studied following ionization by 1.45~$\mu$m pulses. The emission of fluorescence light is a signature of Rydberg atom and ion formation by electron-ion recombination processes during cluster expansion, which radiatively decay on a nanosecond timescale~\cite{ditmire95,schutte14b,schutte15a}. In contrast, the investigation of fast electrons emitted from clusters interacting with intense NIR laser fields provided insight into the dynamics taking place on a timescale of 7 to 1000~fs~\cite{shao96,kumarappan02,springate03,krishnan14}. By using laser pulse durations in the impulsive limit, i.e. so short that the atoms / ions do not move significantly during the laser pulse, it becomes possible to isolate the electron dynamics taking place early during the first few cycles of the laser pulse. This regime has led to the surprising observation that the ion emission was peaked in the direction perpendicular to the laser polarization~\cite{skopalova10}.

Here we report on the first strong-field ionization experiment in clusters using two-cycle laser pulses at 1.8~$\mu$m, meaning that ion motion is effectively frozen. We demonstrate highly efficient electron acceleration, resulting in the observation of keV electrons at much lower laser intensities than in previous NIR experiments~\cite{shao96,kumarappan02,springate03}. A narrow angular distribution shows that electrons are driven out from the cluster by the laser field, and we find signatures of direct electron emission as well as a rescattering plateau. Our results are expected to be relevant to other solid density nanoscale systems interacting with intense few-cycle IR pulses, in which dense plasmas are generated impulsively, including nanostructures, solids and large (bio-)molecules.

\section*{Results}

\subsection*{Electron emission angular distributions}

\begin{figure*}[tb]
 \centering              
  \includegraphics[width=18cm]{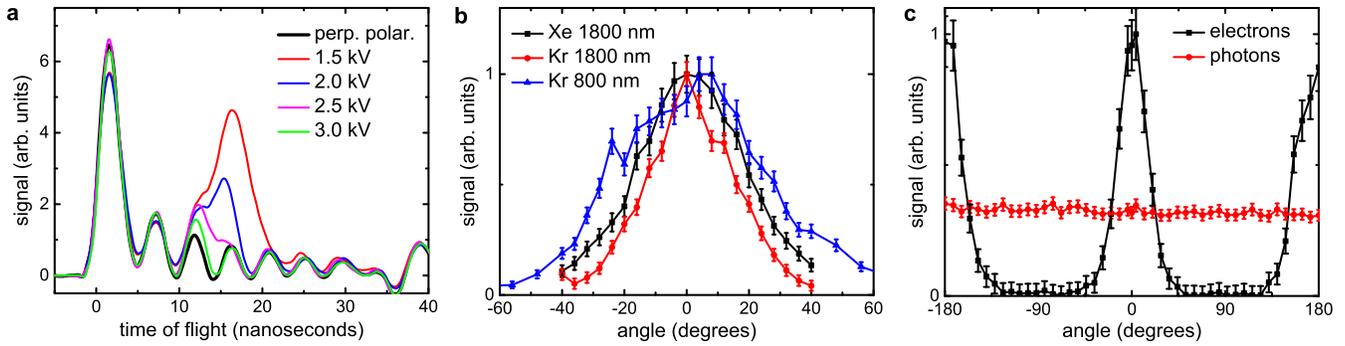}
 \caption{\label{figure1} \textbf{Angular electron emission distributions from clusters.} (\textbf{a}) Electron time-of-flight (TOF) traces measured from Xe clusters ($\langle N\rangle=1\times10^5$~atoms) ionized by 1.8~$\mu$m laser pulses with a duration of 12~fs ($I=1\times 10^{15}$~W/cm$^2$). The black trace was recorded perpendicular to the laser polarization at a retarding voltage of 3000~V, whereas the other traces were measured parallel to the laser polarization at different retarding voltages as indicated. The peak around 1.5~ns is attributed to the emission of photons and exhibits an isotropic behaviour, while the peak between 10 and 20~ns is due to electron emission. Time 0 is defined by the laser-cluster interaction. Some photons appear at negative times due to the finite resolution of the TOF measurement. The oscillations visible after the photon peak are artifacts due to an imperfect impedance matching between the detector and the oscilloscope, which were also observed in previous TOF traces from clusters~\cite{kumarappan02,springate03}, and which do not affect the main conclusions. (\textbf{b}) Angular distributions of electrons emitted from Xe clusters at a retarding voltage of 1000~V (black curve), and from Kr clusters ($\langle N\rangle=23000$~atoms) at a retarding voltage of 500~V (red curve). The different distributions were matched at their peak values. Note that the different retarding voltages only have a small influence on the angular electron emission distribution. For comparison, the angular electron emission distribution from Kr clusters interacting with 13~fs pulses at 800~nm ($I=0.8\times10^{15}$~W/cm$^2$) is shown at a retarding voltage of 100~V (blue curve). This lower voltage was chosen, because the ponderomotive potential of the 800~nm field is significantly smaller, and hence electrons with much lower kinetic energies are observed (cf.~Fig.~\ref{figure2}). (\textbf{c}) Angular emission distribution of electrons (black curve) and photons (integrated signal between 0 and 5~ns, red curve) from Kr clusters over an angular range of 360$^{\circ}$ at a retarding voltage of 1000~V.}
\end{figure*}

In Fig.~\ref{figure1}(a) TOF traces are shown that were recorded from Xe clusters ($\langle N\rangle=1\times 10^5$~atoms) ionized by 1.8~$\mu$m pulses with a duration of 12~fs and an intensity of $1\times 10^{15}$~W/cm$^2$ (see Methods for details). In accordance with previous strong-field ionization experiments on clusters in the NIR regime~\cite{shao96,kumarappan02,springate03}, two peaks are observed. While the first peak at 1.5~ns is only influenced by fluctuations of the experimental parameters, the height of the second peak decreases for increasing retarding voltages. We attribute the first peak to the emission of photons in the ultraviolet and XUV range, in accordance with the experiments performed in Refs.~\cite{kumarappan02,springate03}. These photons are emitted as a consequence of Rydberg atom and ion formation in the expanding cluster and their subsequent decay via fluorescence on a timescale of 1~ns~\cite{ditmire95}. On this long timescale, any information about the laser polarization direction is lost, since during the cluster expansion many collisions between electrons, atoms and ions have taken place. Therefore, the photon emission is isotropic, and the height of this peak does not change for different laser polarizations. The second peak is attributed to electron emission, in agreement with~\cite{shao96,kumarappan02,springate03}. At a retarding voltage of 3~kV (green curve), a fraction of the electron signal remains visible at 12~ns, showing that electrons with energies in excess of 3~keV are emitted from Xe clusters. This is only slightly lower than the maximum electron kinetic energies of 5~keV~\cite{kumarappan02} and 6~keV~\cite{springate03} measured in the NIR regime at much higher intensities of $8\times10^{15}$~W/cm$^2$ and $6\times10^{15}$~W/cm$^2$, respectively. In contrast to the studies in Refs.~\cite{kumarappan02, springate03}, however, ion motion does not play any significant role, since our pulse duration of 12~fs is too short. We attribute the efficient acceleration of electrons as observed in Fig.~\ref{figure1}(a) to the higher ponderomotive potential of the laser field, which is the expected behaviour from studies in atoms reported in Ref.~\cite{colosimo08}. This higher ponderomotive potential partially compensates for the lower laser intensity used in our experiment. 

Angular electron emission distributions from clusters are shown in Fig.~\ref{figure1}(b), and were obtained by rotating the linear laser polarization using an achromatic $\lambda$/2 waveplate. These distributions are peaked along the laser polarization direction. For Xe clusters, the full width at half maximum (FWHM) of the angular distribution is 40$^\circ$ (black curve in Fig.~\ref{figure1}(b)), while it is only about 30$^\circ$ for Kr clusters (red curve in Fig.~\ref{figure1}(b)). This difference could be explained by the larger average cluster size for Xe, which increases the probability of collisions between the escaping electrons and other particles, thus altering the emission direction. Another possible explanation is a different rescattering behaviour in Xe, which is known to be different for different atomic and molecular species~\cite{blaga09}. In comparison, the angular emission distribution of Kr clusters ionized by few-cycle 800~nm pulses is about 55$^\circ$ (blue curve in Fig.~\ref{figure1}(b)). The narrower angular width of the electron emission distribution at 1.8$~\mu$m compared to 800~nm is attributed to the larger ponderomotive energy and the more efficient acceleration of electrons along the laser polarization direction. Fig.~\ref{figure1}(c) visualizes the large contrast in the electron emission in the directions parallel and perpendicular to the laser polarization, whereas photon emission is isotropic. In a strong-field ionization experiment using Ne, which does not form clusters at room temperature, we recorded a FWHM of only 12$^\circ$. This is consistent with earlier measurements on atomic systems, in which a FWHM of about 10$^\circ$ was reported for Ar~\cite{catoire09} and Xe~\cite{chu12} at a wavelength of 2~$\mu$m.

\subsection*{Electron kinetic energy spectra}

\begin{figure}[tb]
 \centering              
  \includegraphics[width=8cm]{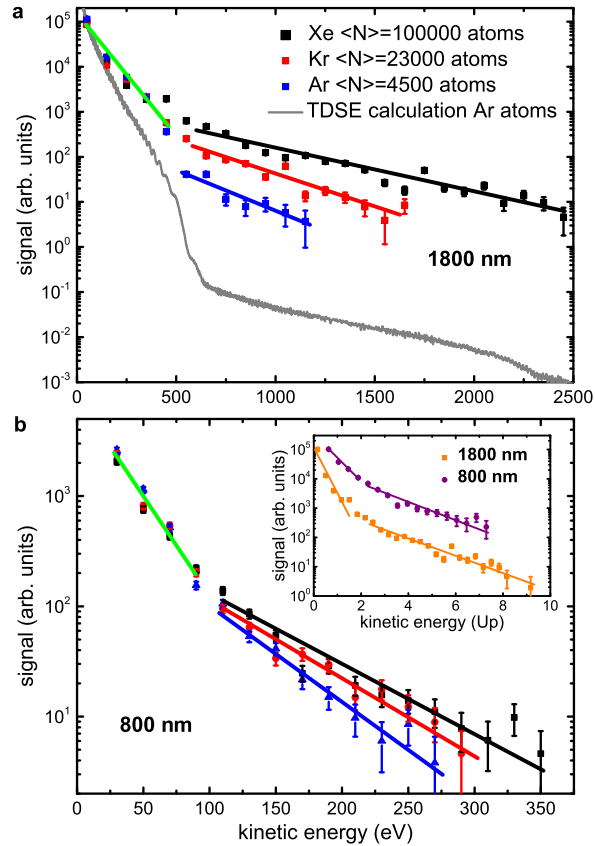}
 \caption{\label{figure2} \textbf{Electron kinetic energy spectra from clusters.} (\textbf{a}) Electron kinetic energy spectra obtained from ionization of Xe, Kr and Ar clusters (with different average sizes as indicated) by 1.8~$\mu$m pulses at an intensity of $1\times10^{15}$~W/cm$^2$ ($U_p\approx 300$~eV) and a pulse duration of 12~fs. The spectra were obtained by measuring electron yields at different retarding voltages and differentiating the results. The lines indicate a different behaviour for electrons $<500$~eV and $>500$~eV. The gray curve is a TDSE calculation for Ar atoms using the same laser parameters as in the experiment. We chose to match the experimental and theoretical data at the first experimental point at 50~eV. (\textbf{b}) For comparison, electron kinetic energy spectra after ionization of Xe, Kr and Ar clusters by 800~nm pulses at an intensity of $0.8\times10^{15}$~W/cm$^2$ ($U_p\approx 50$~eV) are shown. Here the lines indicate a different behaviour for electrons $<100$~eV and $>100$~eV. Note that the absolute values for the data in (\textbf{a}) and (\textbf{b}) are not comparable due to the different experimental conditions. In the inset, the electron spectra from Xe clusters at 1800~nm and 800~nm are directly compared. The data were matched at the first data points.}
\end{figure}

In Fig.~\ref{figure2}(a), electron kinetic energy spectra for clusters consisting of different atomic species and with different cluster sizes are displayed. While the electron yields obtained from Xe, Kr and Ar clusters are similar for kinetic energies up to about 500~eV (as indicated by the green line), they strongly differ at higher kinetic energies (see black, red and blue lines). The fastest electrons are observed from Xe clusters, followed by Kr and Ar clusters. For comparison, we have recorded electron kinetic energy spectra at 800~nm using a pulse duration of 13~fs and an intensity of $0.8\times 10^{15}$~W/cm$^2$, see Fig.~\ref{figure2}(b). Although this intensity is only slightly lower than the intensity of $1\times 10^{15}$~W/cm$^2$ applied at 1.8~$\mu$m, electrons with much lower maximum kinetic energies are observed, in agreement with a recent study at 800~nm using 7~fs pulses~\cite{krishnan14}. 

An electron spectrum from Ar atoms was calculated by solving the time-dependent Schr\"odinger equation (TDSE) for the same parameters as in the experiment (see Methods for details). The result is shown in Fig.~\ref{figure2}(a) (gray curve), where direct electron emission is visible up to 2~$U_P$, which takes place by quasistatic ionization~\cite{corkum89}. Under quasistatic conditions, the yield of direct electrons is given approximately by (atomic units)~\cite{delone99} $Y(E_{kin}) \propto e^{-a E_{kin}}$, where $E_{kin}$ is the photoelectron energy. The slope $a$ is given by $a = ( 2 \omega^2 (2 I_p)^{(3/2)} ) / ( 3 F_0^3 )$, where $I_p$ is the ionization potential, $F_0$ is the peak electric field of the laser, and $\omega$ is the circular frequency. For the Ar curve in Fig~2(a), a quasistatic slope of $\approx 0.1$~Hartree$^{-1}$ is expected. The TDSE direct-electron slope is $\approx 0.5$, indicating that ionization is essentially complete by the time the electric field reaches 60~$\%$ of the nominal peak value. This observation is consistent with the calculated electron dynamics. Interestingly, the experimentally observed slope of $\approx 0.4$ for direct electrons is substantially lower, which is consistent with near-field enhancement that is present in highly ionized clusters~\cite{fennel07b, saalmann08, zherebtsov11}: Enhanced ionization happens earlier within the laser cycle, meaning that the vector potential at the time of ionization is larger than in the case of a bare atom.

In addition, the calculation shows a rescattering plateau at higher kinetic energies, which originates from electrons that gain energy by a laser-driven interaction with ions~\cite{corkum93,walker96,nandor98,colosimo08}. Note that our calculation is expected to slightly overestimate the yield of high energy recollision electrons (see Supplementary Information for more detailed information). This plateau is also observed for clusters, where electrons can be rescattered both by individual ions and by the cluster potential~\cite{saalmann08}. It is striking that the relative yield of rescattered electrons from clusters is by more than two orders of magnitude higher than for atoms. This observation is consistent with the large number of ions in the cluster. We note that the electron scattering mean free path in solid Xe is similar to the cluster radius for electron kinetic energies between 1 and 9~eV~\cite{plenkiewicz86}, and it is expected to increase for larger electron energies. Therefore, electrons that rescatter within the bulk of the cluster may leave the cluster without undergoing additional scattering processes. The rescattering plateau is also visible for 800~nm pulses, but is less pronounced (Fig.~\ref{figure2}(b)), in accordance with results obtained in atoms~\cite{colosimo08}. Also at 800~nm, the signal from clusters drops more slowly towards higher kinetic energies than for atoms, in particular in the region between 2~$U_p$ and 4~$U_p$~\cite{grasbon03}. The inset shown in Fig.~\ref{figure2}(b) displays a comparison between electron spectra recorded for 800~nm and 1800~nm. Similar to atoms~\cite{colosimo08}, direct electron emission exhibits a different slope for the two different wavelengths, whereas the rescattering plateau shows a similar slope in both cases. Rescattering takes place approximately three quarters of a cycle after ionization~\cite{wittmann09}. While ionization can occur early during the laser pulse, the intensity has to be high during the following optical cycle so that the electrons can gain high kinetic energies in rescattering. This limits the emission of fast electrons to a few half cycles~\cite{wittmann09,zherebtsov11}. In the Supplementary Information, a calculated momentum map is presented, showing that rescattered electrons are emitted during at least three half cycles. We note that in addition to direct electron emission and rescattering, thermal electron emission may occur in clusters~\cite{ditmire96}. It is furthermore noted that laser wakefield acceleration can also lead to the generation of fast electrons, but typically requires much higher laser intensities~\cite{modena95}.

\subsection*{Comparison of different clusters and cluster sizes}

\begin{figure}[tb]
\centering              
  \includegraphics[width=8.6cm]{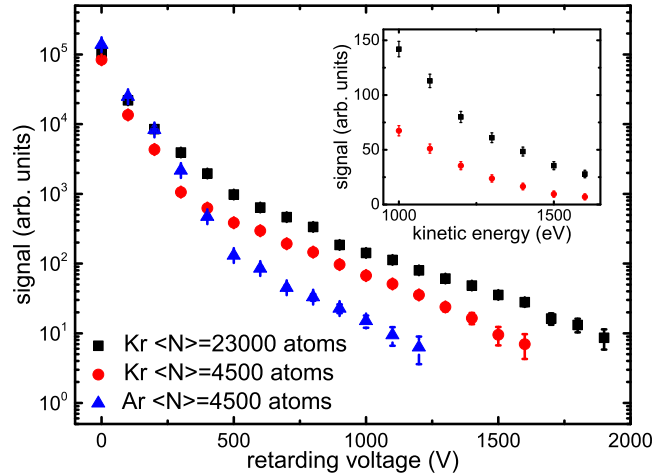}
 \caption{\label{figure3} \textbf{Comparison of different clusters and cluster sizes.} Electron yields for Kr clusters with an average size of $\langle N\rangle=23000$~atoms and $\langle N\rangle=4500$~atoms as well as for Ar clusters with $\langle N\rangle=4500$~atoms recorded at different retarding voltages. The yields for the different clusters and cluster sizes may be influenced by a different number of clusters present in the interaction zone for the different conditions. The inset shows a zoom of the high-energy electrons in a linear scale.}
\end{figure}

In Fig.~\ref{figure3}, the electron yield is shown as a function of the retarding voltage for two different average Kr cluster sizes, and for Ar and Kr clusters with the same average size. Note that in comparison to the differentiated data shown in Fig.~\ref{figure2}(a), the distribution of total yields at different retarding voltages as presented in Fig.~\ref{figure3} is less noisy, but does not represent a kinetic energy spectrum. We find an increased yield of fast electrons for increasing cluster size. At a retarding voltage of 1.5~kV, the electron yield from larger clusters (black squares) is increased by almost a factor 4 in comparison to small clusters (red circles). This clearly indicates a cluster effect, since one would expect that the shape of the electron kinetic energy spectrum remains unchanged, if only the atomic density was varied. The quantitative differences in the cluster-size dependent spectra can be explained by different widths of the scattering potential that are expected for clusters with different sizes~\cite{saalmann08}. The different behaviour of Ar and Kr clusters with the same average size, as observed in Fig.~\ref{figure3}, show that also the atomic structure has an influence on the rescattering plateau. Future time-resolved experiments could give insights in the contributions of the atomic and the cluster potential on rescattering processes.

\section*{Discussion}

The high relative yield of fast electrons from clusters presented in this work may be due to near-field enhancement effects that were previously studied for clusters~\cite{fennel07b, saalmann08} and nanoparticles~\cite{zherebtsov11}. Model calculations could give insights into the underlying physical mechanisms. However, such calculations for our experimental parameters that would have to take into account both rescattering processes at the potential of single ions and at the potential of the cluster as a whole are at the limits of what is currently feasible. We hope that our work will stimulate the theoretical advances required to allow realistic modelling of such experiments. The understanding of the involved processes could further be improved by performing time-resolved experiments with attosecond resolution.
																												
In summary, we have observed highly efficient acceleration of electrons to multi-keV levels from clusters in strong two-cycle laser fields at 1.8~$\mu$m that shows an anisotropic distribution peaked along the laser polarization direction. We found clear signatures of direct and rescattering processes with a much higher probability compared to atoms. This scheme may therefore lead to efficient HHG and even attosecond pulse generation from clusters in the water window. Scaling the ionization wavelength to the multi-terahertz regime~\cite{schubert14,hohenleutner15} is expected to result in a highly directional current of energetic electrons within a few femtoseconds. The presented results are pertinent to the laser driven electron dynamics whenever an intense few-cycle laser pulse interacts with condensed phase matter.

\section*{Methods}

\textbf{Experiment.}

For the generation of 1.8~$\mu$m pulses, linearly polarized laser pulses at 800~nm with an energy of 8~mJ and a duration of 30~fs were derived from a Ti:sapphire amplifier system (1~kHz repetition rate), and coupled into an optical parametric amplifier. The linearly polarized output idler pulses at 1.8~$\mu$m were spectrally broadened by focusing them into a differentially pumped hollow-core fiber filled with Ar~\cite{austin16}. Temporal compression of these pulses was achieved by a combination of a 1~mm BK7 window and an achromatic $\lambda/2$ waveplate consisting of a 1.5~mm fused silica plate and a 1.2~mm MgF$_2$ plate. In addition, the waveplate was used to rotate the linear polarization. The pulse duration was measured before and after each experimental run making use of spatially-encoded arrangement filter-based spectral phase interferometry for direct electric field reconstruction (SEA-F-SPIDER)~\cite{witting09,witting12}. The laser path from the output of the fiber to the experimental chamber through air and through the vacuum window slightly stretched the pulse, resulting in a typical pulse duration of 12~fs FWHM. After transmission and reflection losses, a typical pulse energy of 0.45~mJ was applied in the experiment. We used a second hollow-core fiber filled with Ar in order to spectrally broaden the 800~nm pulses. Temporal compression of these pulses was achieved by chirped mirrors and an achromatic waveplate (that was also used for polarization control), giving a pulse duration of 13~fs.

The compressed laser pulses were coupled into a vacuum chamber through a 1~mm thick CaF$_2$ window, and focused into the interaction zone by a spherical silver mirror with a focal length of 15~cm at near-normal incidence. In the interaction zone, the laser beam and a cluster beam that was generated by a piezoelectric valve intersected at right angles. The average cluster size was controlled by varying the backing pressure between 1 and 7~bar, and estimated according to the Hagena scaling law~\cite{hagena72}.  Under these conditions clustering is known to be highly efficient and to lead to clusters in the size range from 10$^3$ to $10^5$ atoms, corresponding to average cluster radii between about 3 and 15~nm. We confirmed efficient cluster generation in an independent experiment, where ions with substantial kinetic energies were observed as a result of cluster explosions. A molecular beam skimmer with an orifice diameter of 0.5~mm was used to select the central part of the cluster beam and to maintain a low pressure in the experimental chamber. 

A TOF spectrometer with an acceptance angle $<3^\circ$~\cite{hemmers98} was used for the simultaneous detection of electrons and photons. A retarding voltage could be applied to the flight tube such that only electrons with kinetic energies above the retarding potential were detected. Kinetic energy spectra were obtained by measuring the electron yields at different retarding voltages and differentiating the results, similarly to previous studies of fast electrons emitted from clusters~\cite{shao96,kumarappan02,springate03}. An achromatic $\lambda/2$ waveplate was used to rotate the linear laser polarization axis with respect to the TOF axis. 

\textbf{Calculations.}

We have calculated a photoelectron spectrum for Ar atoms using TDSE~\cite{patchkovskii16,morales16} for the same laser parameters as in the experiment (gray curve in Fig.~\ref{figure2}(a)). One-electron TDSE calculations~\cite{patchkovskii16} used an effective potential designed to reproduce the energies of the low-lying states of an argon atom~\cite{morales16}. TDSE was solved in a velocity-gauge dipole approximation on a spherical grid. Angular momenta up to $L=500$ were included. A uniform radial grid (grid spacing $0.1$~Bohr) extended to $450$~Bohr from the origin, with a 33-Bohr wide absorbing boundary~\cite{Manolopoulos02} starting at 417~Bohr. A truncated Gaussian~\cite{patchkovskii16} laser pulse with a carrier wavelength of 1800~nm and a FWHM of 12~fs was centered at 36.3~fs. The envelope was smoothly switched to zero between 24 and 36~fs from the envelope peak. A sine carrier envelope phase was used, yielding six half-cycles (three on each side of the envelope peak), where the instantaneous field intensity exceeds 30~$\%$ of the nominal peak intensity. The electric field of the laser pulse is shown in Fig.~1 of the Supplementary Information. The laser field was polarized along the $Y$ direction. The simulation continued up to 72.6~fs, using time steps of $\approx$ 4.84~zs, corresponding to a total of $1.5\times 10^7$ time steps. The atomic $3p_y$ function ($m=0$ along the laser field polarization direction) was used as the initial state for solving the TDSE. More details about the calculations can be found in the Supplementary Information.


\section*{Acknowledgments}

We thank Mathias Arbeiter and Thomas Fennel for fruitful discussions. B.S. is grateful for funding from the DFG via a research fellowship. Funding from the following grants is also gratefully acknowledged: EPSRC (EP/I032517/1) and the ERC (ASTEX project 290467) and EPSRC/DSTL MURI grant EP/N018680/1.

\section*{Author contributions statement}

B.S. and P.Y. carried out the experiments, and B.S. analysed the experimental data. S.P. performed the theoretical calculations. D.R.A., C.B., P.Y. and C.S. took care of laser alignment, compression and frequency conversion. J.P.M. supervised the project. All authors discussed the results. B.S. and J.P.M. wrote the manuscript, to which the other authors contributed.

\section*{Additional information}

\textbf{Competing financial interests:} The authors declare no competing financial interests. 

\end{document}